\documentclass[aps,preprint,eqsecnum,prl]{revtex4}
\usepackage{epsfig}

\newcommand{\beq}{\begin{equation}}
\newcommand{\eeq}{\end{equation}}

\newcommand{\vev}{{\langle\, \phi\, \rangle}}
\newcommand{\vevphi}{{\langle\ \phi\ (\varphi)\ \rangle}}
\newcommand{\vevphial}{{\langle\ \phi\ (\varphi, \alpha)\ \rangle}}

\newcommand{\remove}[1]{}

\newcommand {\comment}[1]{}             

\begin{document}


\title{A Topological Criterion for Alice Strings}

\author{Katherine M.~Benson}
\email{benson@physics.emory.edu}
\affiliation{
Department of Physics,
Emory University,  400 Dowman Drive, Suite N202,Atlanta, GA\ 30322}
\author{Tom Imbo}\email{imbo@uic.edu}
\affiliation{Department of Physics,
University of Illinois at Chicago,
845 W. Taylor St,  m/c 273,
Chicago, IL 60607-7059}

\date{\today}

\begin{abstract}
Symmetry breaking can produce ``Alice'' strings, which alter scattered
charges and carry monopole number and charge when twisted into loops.
Alice behavior arises algebraically, when a string's untraced Wilson
loop obstructs unbroken symmetries --- a fragile criterion. We give a
topological criterion, compelling Alice behavior or deforming it
away. Our criterion, that $\pi_o(H)$ acts nontrivially on $\pi_1(H)$,
links topologically Alice strings to topological monopoles. We twist
Alice loops to form monopoles, and find nematic and $^3$He-A Alice
strings are topologically Alice, and carry fundamental monopole charge when
twisted into loops.

\end{abstract}

\maketitle



\section{Introduction and Our Criterion}

Among the defects created when gauged symmetries break down are Alice
strings.\cite{oldAlice,stringzm,newAlice,blp} Alice strings obstruct
the global extension of unbroken symmetries, making them multivalued
when parallel transported around the string. This algebraic
obstruction has two prominent physical consequences. First, it
produces nonconservation of associated charges, when Aharonov-Bohm
scattered around the string. Second, it induces monopoles, as twisted
loops of Alice string.

These Alice features arise due to gauge flux on the string's core.
The gauge flux generates the condensate winding, while acting on
asymptotic particles through the Wilson line $U(\varphi)$. This action
fixes particles' Aharonov-Bohm scattering around the string, to one
changing both charge and monopole number. Loops of string, which
leave charge and monopole number well-defined asymptotically, thus
support deposited unlocalizable charge (``Cheshire charge'') and
deposited monopole number. \cite{oldAlice,stringzm} Alice loops carry
this deposited monopole number by twisting, \cite{blp} as we probe
further below.

Alice strings in condensed matter systems are global, not gauged,
defects. They have no gauge flux to fix their Aharonov-Bohm
scattering, and guarantee altered charge and monopole number upon
string traversal. However, their Aharonov-Bohm scattering was
considered in \cite{ABrefs, global}; with \cite{global} showing that
global Alice strings generically share all Alice behaviors.
They alter both charge and monopole number on string
traversal, with twisted loops supporting both Cheshire charge and
deposited monopole number.

The criterion for Alice string formation was first stated
algebraically, in terms of the string's untraced Wilson loop
$U(2\pi)$.  When $U(2\pi)$ fails to commute with an unbroken symmetry
$h$, the symmetry cannot be globally extended; when it fails to
commute with unbroken generator $T_h$, the associated charge is
nonconserved. Thus Alice strings arise when $U(2\pi)$ lies outside the
center of the unbroken symmetry group $H$. As noted in
\cite{stringzm}, this is an inherently nontopological criterion, as
topologically equivalent choices for $U(2\pi)$ can commute with
different subgroups of $H$. Thus
emergence of Alice behavior appeared a dynamical question. Steps
toward topologizing this criterion came in \cite{blp}. They noted that
when {\bf all} topologically equivalent choices for $U(2\pi)$ lie
outside the center of $H$, Alice strings must form. Equivalently,
topological Alice strings form when the fiber bundle of $H$ parallel
transported around the string is nontrivializable. Both criteria,
while accurate, seem difficult to apply.

We here establish an easily applied topological criterion which states
when Alice strings {\bf must} form.  We take $G$ to be the simply
connected cover of the initial symmetry --- a connected Lie group ---
and $H\subset G$ its unbroken subgroup.
A topologically stable string has flux
$U(2\pi)$ in a disconnected component of $H$, with
 topology determined by $\pi_o(H)$. Similarly,  the
topology of the monopole is given by $\pi_1(H)$, describing loops
$h(\alpha)$ of different winding in $H$. By taking seriously the
change in  monopole number in circumnavigating the Alice
string, we construct our criterion. Note that, in Aharonov-Bohm
scattering around the string, the monopole $h(\alpha)$ is conjugated
by the string's Wilson loop $U(2\pi)$:
$$h(\alpha) \rightarrow \tilde{h} (\alpha) = U(2\pi)\ h(\alpha)\
U^{-1}(2\pi)\ \ .$$ Monopole number changes if $\tilde{h}(\alpha)$ and
$h(\alpha)$ are topologically distinct loops. We
represent this transformation topologically, as $\pi_o(H)$ acting
naturally on $\pi_1(H)$ by conjugation. Topological Alice strings form if
that action is nontrivial: that is, if, for $h_o$ a representative
element of $\pi_o(H)$ and $h(\alpha)$ a representative loop in
$\pi_1(H)$,
$$ \tilde{h} (\alpha) = h_o\ h(\alpha)\ h_o^{-1} \ \ \not\sim \ \
h(\alpha)\ \ .$$
A string with untraced Wilson loop $U(2\pi) \ \sim \ h_o$
meeting this criterion is topologically guaranteed to change monopole
number; we dub it a topologically Alice string.

This criterion captures physical Alice behavior, is easily applied,
and is topological. Its result, for any chosen $ h_o$ and $h(\alpha)$, remains invariant under deformations
of either flux $U(2\pi) = h_o$ or monopole loop
$h(\alpha)$. By construction, the strings are topologically Alice if
monopoles change topologically in traversing them. Of
course, monopoles $h(\alpha)$ change because of algebraic Alice
behavior: loop $h(\alpha) = e^{i\alpha T_h}$ alters only if its
generator $T_h$ alters; that is, if $U(2\pi)$ fails to commute with
generator $T_h$. This algebraic noncommutation creates the standard
Alice constellation of behaviors: multivalued symmetry,
charge-violating Aharonov-Bohm scattering, Cheshire charge on Alice
loops. It is captured by our criterion {\bf only} when altering
generators  alters the topology of the loops they
generate. This misses some Alice phenomena --- particularly in models
with poorly distinguished loops, when $\pi_1(H) = 0$ (and all loops are trivial) or
$\pi_1(H) = Z_2$ (and all nontrivial loops, including a loop and its inverse, are identified). We claim that Alice behavior in these models is
not robust topologically; that is, continuous deformation of such
strings removes their Alice behavior. In such cases, persistence of
Alice behavior can arise only from dynamical arguments, favoring
Alice strings over non-Alice strings of the same
winding. Dynamically stabilized features remain interesting --- for example, nontopological defects including embedded, semilocal, or
electroweak strings. \cite{semew} However,  we seek here for Alice behavior 
the more robust motivation of topological imperative.

Note that this topological imperative comes at a cost: to fulfill our
topological criterion, of $\pi_o(H)$ acting nontrivially on
$\pi_1(H)$, $\pi_1(H)$ itself {\bf must be nontrivial}.  That is, only
in a theory with topological strings and monopoles --- and more pointedly, monopoles
topologically distinct from antimonopoles --- do topologically Alice
strings arise.

Our topological criterion for Alice strings ensures that twisted loops
carry monopole charge, as we see by explicit construction below.
We note that topological arguments only indicate that
{\bf deposited} monopole charge can be carried by twisted Alice
loops. In many models, Alice strings interchange monopoles with
antimonopoles, depositing monopole charge in units of 2.  Whether
fundamental monopoles, or only those with even charge, are deformable
to twisted loops of Alice string is
model-dependent. We see both possibilities arise in
\cite{loopmon}.

We present our results as follows.
We first show
that Alice strings failing our topological test have topologically unstable Alice
features; that is, their Alice
behavior can be deformed away.  We then
show, by construction, that twisted Alice strings carry monopole
charge.
We argue topologically, using our criterion to display a
twisted Alice loop, carrying monopole number deposited in the
monopole scattering $h(\alpha) \rightarrow \tilde{h}(\alpha)$. We then
illustrate our criterion and Alice loop twisting to form monopoles,
for key models: the Schwarz Alice string,
coinciding with the Alice string of liquid crystals and of non-chiral
Bose condensates \cite{oldAlice, leonvol} and a nontopologically Alice
string introduced in
\cite{stringzm}.
Key points include, for the nontopological model, a focus on how Alice
candidates may fail our criterion; and for the Schwarz string,
analysis of the monopole charge carried by twisted Alice loops. In
both  Schwarz and $^3$He-A Alice models, Alice string scattering
changes monopole number by 2, yet
twisted Alice loops carry a single fundamental unit of monopole
charge. Thus, for Alice strings in condensed matter, even the
fundamental monopole can be, in fact, a
twisted Alice loop. In contrast, model Alice loops discussed
elsewhere  support only even, not fundamental, monopole
charge.\cite{loopmon} In this paper we elucidate our topological criterion and its generic consequences for twisted Alice loops; we elaborate consequences of the criterion for prominent condensed matter models in a companion paper \cite{loopmon}.


\comment{
We note that while most proposed Alice strings have $Z_2$ Alice
behavior, with generators, charges, and monopole number reversing sign
on circumnavigating the string, our topological analysis makes no such
assumption. Our criterion reveals any multivaluedness induced by the
Alice string, {\bf if} that multivaluedness, and associated Alice
phenomena, are topologically compelled.}

\section{\label{nontop} Failure means Topological Instability}

Consider an Alice string with Wilson loop $U(2\pi) = h_o$. The Alice
string fails our topological criterion if, for a nontrivial loop
$h(\alpha)\ \in\ H$ describing a monopole, the parallel transported
monopole is homotopic to  the original; that
is,
\beq\tilde{h} (\alpha) = h_o\ h(\alpha)\ h_o^{-1} \ \ \sim \ \
h(\alpha)\ \ .\label{crit}\eeq
This occurs only if there exists some continuous map $f(x)$ deforming $\tilde{h} (\alpha)$ to $h(\alpha)$; that is
$$f(x) \ :\ \tilde{h} (\alpha) \ \ \rightarrow\ \ \left\{
\ \begin{array}{cc}
\tilde{h} (\alpha) & \mbox{when $x=0$} \\ h(\alpha) &\mbox{when $x=1 $}
\end{array}\ \ . \right.  $$
Note that the map $f(x)$
relates nontrivial loops in $H$, with basepoint $\alpha = 0$ fixed at the
group identity.  We write it  as a continuously varying group element $f(x)$ acting on $\tilde{h} (\alpha)$ by conjugation,
$$f(x) \ :\ \tilde{h} (\alpha) \ \ \rightarrow\ \ \ f(x)\
\tilde{h}(\alpha)\ f(x)^{-1}\ \ .$$ Without loss of generality we
take $f(0) = 1\mkern-6mu 1$.

Now consider the continuous map
$ {h'}_o \ (x)= f(x)\ h_o\ ,$ where $f(x)$ acts on $h_o$ by left
multiplication. This interpolates between the Alice string's Wilson
loop $h_o$, and group element $ {h'}_o (1)$ in the same disconnected
component of H.  $ {h'}_o  (1)$ thus defines a topologically
equivalent string, with Wilson loop $U(2\pi) ={h'}_o  (1)$. In circumnavigating this deformed string, the
original monopole $h(\alpha)$ is unchanged: it goes to
\begin{eqnarray*}\tilde{h'} (\alpha) &=& {h'}_o \, (1)\ h(\alpha)\ {h'}_o^{-1}  \, (1)= f(1)\ h_o\
h(\alpha)\ {h}_o^{-1}\ f(1)^{-1}\\ &=& f(1)\ \tilde{h}(\alpha)\
f(1)^{-1} = h(\alpha)\ \ , \end{eqnarray*}
by construction of $f(x)$. Thus the
monopole loop $h(\alpha)$ remains identical on circumnavigating the
string. Choosing as our nontrivial loop $h(\alpha) = e^{i\alpha
T_h}$, the loop (at each value of $\alpha$) remains unchanged only if
the generator $T_h$ remains unchanged. Thus by continuously deforming
our Alice string's flux from $h_o$ to ${h'}_o  (1)$, we have obtained a
string flux $U(2\pi) = {h'}_o  (1)$ which commutes with all generators;
that is, we have removed all Alice behavior of the string. This renders Alice
behavior for  strings failing our criterion  topologically
unstable; it can be deformed away, and stabilized only in
dynamical, model-dependent ways.

\section{\label{monopoles}Twisted Alice loops are monopoles}

Monopoles lie on the vacuum manifold at spatial infinity, with
topology given by $\pi_2(G/H)$. We here show that a twisted
topologically Alice loop is necessarily a topological monopole; that is,
an infinite sphere enclosing it has nontrivial $\pi_2(G/H)$.

First, we construct a sensible twisted Alice loop.

Recall that our Alice string has a condensate $\vev$ which winds
asymptotically over the vacuum manifold $G/H$ according to
$$ \vevphi = U(\varphi)\ \vev_o\ ,
$$
where the Wilson line $U(\varphi)$  acts on the vev $\vev_o$ according to its group representation.  $U(\varphi)$
 varies  continuously over $G$ for $0<\varphi<2\pi$, and connects the identity at $\varphi =0$ to a distinct Wilson loop $U(2\pi)$ in $H$. The string
is topological when $U(2\pi) = h_o$ lies in a disconnected component of $H$,
with nontrivial $\pi_o(H)$, and  is topologically Alice when it meets
our criterion (\ref{crit}).

Now twist the Alice string: continuously rotate its Wilson line within $G$ by  the angle-dependent $H$-group rotation $h^{-1}(\alpha)$:
$$U(\varphi, \alpha) = h^{-1}(\alpha)\ U(\varphi)\ h(\alpha)\ \ ,$$
as shown in Figure 1a. This, of course, rotates our condensate among the degenerate vacua on $G/H$:
$$ \vevphial = U(\varphi, \alpha)\ \vev_o\ \ .
$$

Under what conditions may we identify string ends at $\alpha = 0$ and
$\alpha = 2\pi$ to form a string loop, as pictured in Figure 1b?
First, we require the string configurations to match at the
junction. This is assured if $h(2\pi) = h(0)$, that is, if $h(\alpha)$
is a loop. Second, the twisted condensate $\vevphial$ must be single-valued. Note that the Wilson line $U(\varphi, \alpha)$ itself need not be single-valued: indeed, for a monopole configuration, $U(\varphi, \alpha)$ interpolates from the identity at $\varphi =  0$ to a nontrivial loop in $H$ at $\varphi = 2\pi$.

First we check singlevaluedness of $\vevphial$ at the loop's origin. Here $\varphi = 0$ (or $2\pi$) while
$\alpha$ is indeterminate. Note that $ U(0, \alpha)$ is the identity, manifestly single-valued. At $\varphi = 2\pi,$
$$U(2\pi, \alpha) =  h^{-1}(\alpha)\ U(2\pi)\ h(\alpha)\ \ .$$
This  generally does vary with $\alpha$; however,  it is a loop in $H$, with basepoint $U(2\pi) = h_o \in H$. It thus leaves the condensate invariant, assuming the single value $\vev_o$  at loop origin.

Elsewhere, we need only show first, that $\langle\ \phi\ (\varphi,
\alpha + 2\pi)\ \rangle = \vevphial$; and second, that $\langle\ \phi\
(\varphi + 2\pi, \alpha)\ \rangle = \vevphial$. The first is trivial:
since $h(\alpha)$ is a loop, $h(\alpha) = h(\alpha + 2\pi)$ and both
$U(\varphi, \alpha)$ and $\vevphial$ are single-valued in $\alpha$.

To show singlevaluedness in $\varphi$, let us, without loss of generality, diagonalize our string Wilson line $U(\varphi)$, taking it to be generated by a fixed generator so that
$U(\varphi +  2\pi ) = U(\varphi)\ U(2\pi)\ .$
Then our twisted Wilson line obeys
$$U(\varphi +  2\pi, \alpha) =  h^{-1}(\alpha)\,\, U(\varphi)\,\, U(2\pi)\,\, h(\alpha)
= U(\varphi, \alpha)\,\, U( 2\pi, \alpha)\  .$$
As noted above, $U( 2\pi, \alpha)$ is a loop in $H$, leaving $\vev_o$ invariant. Thus
$$\langle\ \phi\ (\varphi, \alpha + 2\pi)\ \rangle = \vevphial =  U(\varphi, \alpha)\ \vev_o\ \ $$
and our twisted Alice loop is fully single-valued.

By the exact sequence for $\pi_2 (G/H)$, our twisted Alice loop is a
monopole when $U(\varphi, \alpha)$ interpolates between an element of $H$ at
$\varphi = 0$ and a nontrivial loop in $H$ at $\varphi = 2\pi$. For convenience, right multiply  $U(\varphi, \alpha)$ by $h_o^{-1}$:
\beq U(\varphi, \alpha) = h^{-1}(\alpha)\ U(\varphi)\ h(\alpha)\ h_o^{-1}\ \
.\label{twistmon}\eeq
Since $h_o^{-1} \in H$, this right multiplication does not change the physical condensate $\vevphial$. However, it makes the topology of $U(\varphi,\alpha)$  clear, for
$$U(\varphi,\alpha) = \left\{\ \begin{array}{ll} h_o^{-1}& \mbox{for}\ \varphi = 0\\ h^{-1}(\alpha)\ \tilde{h}(\alpha)& \mbox{for}\ \varphi = 2\pi\end{array}\right. \ \ .$$
By definition, if the string is topologically Alice,
$\tilde{h} (\alpha) \ \not\sim \
h(\alpha)$ so that $ h^{-1}(\alpha)\ \tilde{h}(\alpha)$ is a nontrivial loop in $H$, and  the twisted Alice loop carries nontrivial monopole charge. If the string is {\em not} topologically Alice, the loop $ h^{-1}(\alpha)\ \tilde{h}(\alpha)$ is trivial in $H$ and the twisted Alice loop carries no monopole charge.

Thus twisted Alice loops carry monopole charge if and only if they
obey our topological Alice criterion. We note that the
monopole charge displayed , with winding $ h^{-1}(\alpha)\ \tilde{h}(\alpha)$, is
exactly that deposited on an initially untwisted Alice loop,
when a monopole of winding $h^{-1}(\alpha)$ circumnavigates the string
and emerges with winding $\tilde{h}^{-1}(\alpha)$. \cite{blp} (The {\em inverse} twisted
Alice loop, generated by $U^{-1}\ (\varphi, \alpha)$, instead carries
monopole charge $h(\alpha) \
\tilde{h}^{-1}(\alpha)$, deposited in the monopole circumnavigation
$h(\alpha) \ \rightarrow \ \tilde{h}(\alpha)$.)

We note that our final map $U(2\pi,\alpha)$ for the twisted
Alice loop coincides with the flux loop (4.1) and paths $C'_\alpha$
defined in \cite{blp}. They show that this map coincides with the
Lubkin classification of monopole charge for the twisted Alice loop.
\comment{which considers only flux variation over a 1-parameter loop of closed
paths on the sphere at spatial infinity, enclosing the Alice
loop.} This reinforces our classification, as  identifying
twisted topologically Alice loops with physical gauged magnetic monopoles.

\section{\label{model:canon}The Schwarz, or Nematic,  Alice String}

The simplest Alice string is that of  Schwarz,
\cite{oldAlice} whose symmetry structure coincides with
Alice strings in nematic liquid crystals and
non-chiral Bose condensates. \cite{leonvol}
Here $G$ is $SO(3)$, with
Higgs $\phi$ transforming in the adjoint.
$\phi$ develops the vev
$\vev = {\rm diag}\ (1,1,-2)\ ,
$
breaking $SO(3)$ to the residual symmetry $H = O(2)$, containing  z-rotations $R_z\,(\alpha)$ and the discrete symmetry element
$ h_o = R_x\, (\pi) = {\rm diag}\ (1, -1, -1)\ .$
Here $\pi_o (H) = Z_2$ and $\pi_1(H) = Z$ so we have topological strings and monopoles. The Alice string has Wilson line
$ U(\varphi) = R_x\, (\varphi/2)$
with $U(2\pi) = h_o$. $U(2\pi)$ fails to commute
with unbroken symmetry generator $T_z$; in fact, on parallel
transport around the string,
\beq T_z \rightarrow U\ (2\pi) \ \ T_z \ \ U^{-1}\ (2\pi) = -T_z \ \ .
\label{canAliceT}\eeq

This Schwarz Alice string meets our topological criterion, of changing
topological monopole charge on circumnavigation. By the exact
sequence for $\pi_2 (G/H)$, topological monopoles are associated with
nontrivial loops in $O(2)$ which can be unwound in $SO(3)$. Since only
even winding loops in $O(2)$ can be unwound in $SO(3)$, the
fundamental monopole in this canonical Alice model has a loop in
$O(2)$ of winding 2.

To apply our topological criterion, we
represent the string by $h_o$, a nontrivial element of
$\pi_o (H)$, and the fundamental monopole by $h(\alpha) = R_z(2\alpha)$, a winding 2 element of $\pi_1(H)$. This gives
$$\tilde{h} (\alpha) = h_o\ h(\alpha)\ h_o^{-1}  = h^{-1} \, (\alpha)\ \ ,$$
from equation (\ref{canAliceT}). Note that $h^{-1} \, (\alpha)$ has $O(2)$ winding -2, topologically distinct from $h(\alpha)$ of $O(2)$ winding 2. Thus
$ \tilde{h} (\alpha)\ \not\sim \
h(\alpha)$ and our topological criterion is met.

We now construct a monopole as a twisted Alice loop. From Eq.~(\ref{twistmon}), the twisted Wilson line
$$ U(\varphi, \alpha) = h^{-1}(\alpha/2)\ U(\varphi)\ h(\alpha/2)\ h_o^{-1}\ \
$$
gives an Alice loop with single-valued condensate. (We take
$h(\alpha/2)$ because $h$ need only  be single-valued in
$\alpha$, and $h(\alpha/2)$, the winding 1 loop in $O(2)$, first
achieves this.)  $U(\varphi, \alpha)$ interpolates between
$ h_o^{-1}$ at $\varphi = 0$ and $ h^{-1}(\alpha)$ at $\varphi =
2\pi$. It is thus the fundamental antimonopole in the model,
winding $-2$ in $O(2)$. The inverse twisted Alice loop, with Wilson
line $ U^{-1}(\varphi, \alpha)$, creates the fundamental monopole.

Similarly, twisted Alice loops in $^3$He-A and amorphous chiral superconductors support fundamental
monopole charge, as again only even $U(1)$ loops unwind inside
$SO(3)$.\cite{loopmon} Embedding $H = O(2)$ in a different $G$,
however, can result in Alice loops unable to support fundamental
monopole charge. \cite{loopmon}

\section{\label{model:nontop}A Nontopologically Alice string}

Consider the nontopologically Alice string introduced
in \cite{stringzm}: a Higgs $\phi$, transforming
in the adjoint under $G=SO(6)$, acquires the vev $\vev
= {\rm diag}(1^3,-1^3)$. This
condensate leaves unbroken an $SO(3)\times SO(3)$ subgroup of $SO(6)$
and a discrete $Z_2$ transformation $h_1=-1\mkern-6mu 1_6$, so
$H=SO(3)\times SO(3)\times Z_2$.  Here $\pi_o(H)=Z_2$ and $\pi_1(H) =
Z_2 \times Z_2$, so topological strings and monopoles form,
with monopoles and antimonopoles identified.   Alice characterististics of the string
depend on $U(2\pi)$.  For $U(2\pi) = h_1$,
 all unbroken generators $T_{ij}$ are single-valued under parallel transport around the string,
and the string is not Alice. However, for the topologically equivalent choice
$U(2\pi) = {\rm diag} (1^2, (-1)^4) = - R_{12}(\pi) $,
the
string is Alice, making  generators $T_{13}$ and $T_{23}$ of rotations in
the $13$- and $23$-planes double-valued.
Since this Alice behavior is removable by deforming to the
topologically equivalent string with $U(2\pi) = h_1$, it must be
nontopological. However, it is instructive to see how the two strings
fail our topological criterion. Taking as our  nontrivial monopole loop $h(\alpha) = R_{13}\ (\alpha)$, we obtain, for $h_o = h_1$,
$\tilde{h} (\alpha) = h_o\ h(\alpha)\ h_o^{-1} = h \, (\alpha)\ \ .$
That is, our topological criterion fails, as the monopole remains
unchanged in circumnavigating the Alice string. Instead, for
$h_o =  - R_{12}(\pi), $
$\tilde{h} (\alpha) = h_o\ h(\alpha)\ h_o^{-1} = h^{-1} \, (\alpha)\
\ ,$ since $T_{13}\rightarrow -\ T_{13}$. Here a monopole transforms into an
antimonopole on traversing the string. {\bf However,} that
transformation is nontopological, as monopoles and antimonopoles are
topologically equivalent. So, despite algebraic Alice
behavior, this string is not topologically Alice. Loops
$h(\alpha)$ alter on string traversals, but in a
topologically trivial way.

\section{\label{conclusions}Conclusions}

We have established a topological criterion for strings to display
Alice behavior. This criterion, that $\pi_o(H)$ acts nontrivially on
$\pi_1(H)$, depends only on the residual symmetry group $H$. Alice
strings must form in models obeying this criterion, while Alice
behavior can be deformed away for strings failing the criterion. Particularly, the criterion requires that topological monopoles always accompany Alice strings; and furthermore, that Alice strings alter the topological
charge of monopoles that circumnavigate them. We construct
monopoles as twisted loops of Alice string, and show that such twisted
loops can always support deposited monopole charge. Whether twisted
Alice loops can support fundamental monopole charge  depends on the
symmetry-breaking pattern more closely, as we discuss in \cite{loopmon}. For the Schwarz  Alice string, we
construct a twisted Alice loop that supports fundamental monopole charge.

\begin{acknowledgments}
Early stages of this work were supported by NSF grant PHY-9631182 and
by the University Research Committee of Emory University. KB thanks
the KITP (under NSF grant PHY99-07949)  for hospitality during the writing of
this paper.
\end{acknowledgments}

\newpage

$$\begin{array}{c}
\epsfig{file=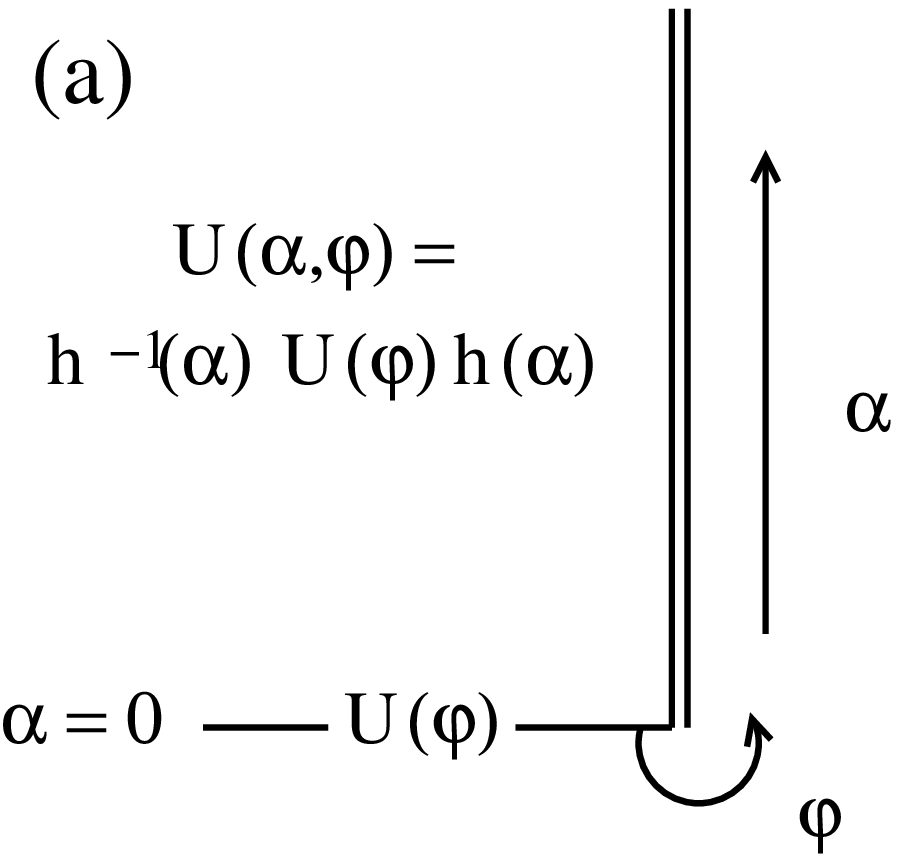, width = 4.0cm}\quad\quad\epsfig{file=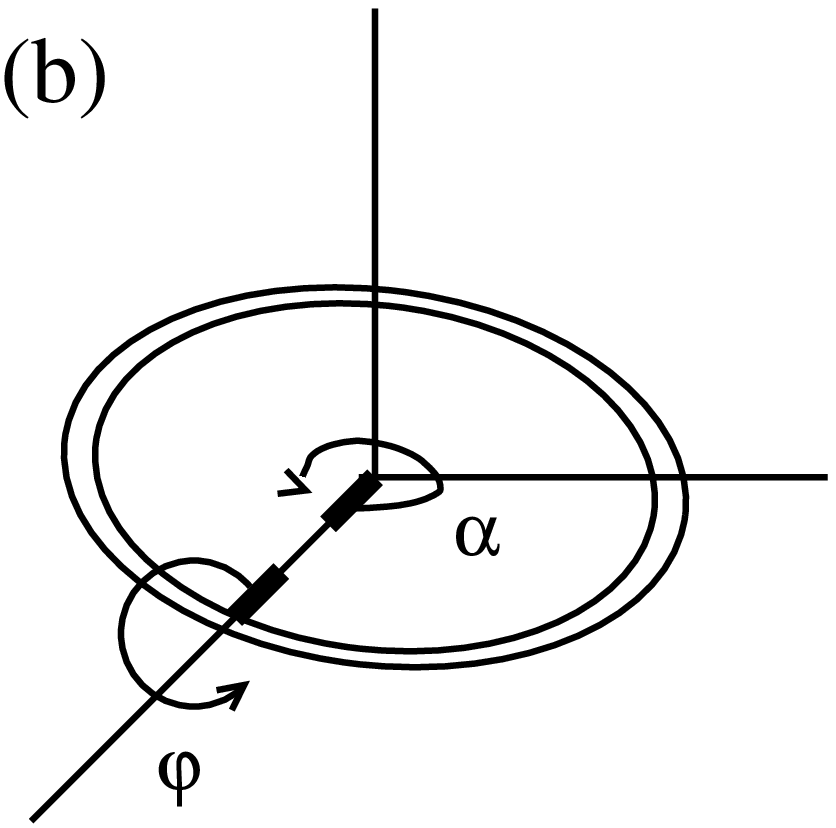, width = 4.0cm}\\
\parbox{8.6cm}{FIG. 1: a) A twisted Alice string. b) Identifying twisted Alice string ends to form a twisted Alice loop.}
\end{array}$$

\end{document}